\documentstyle[psfig,epsfig,aps,aps12]{revtex}
\def\be{\begin{equation}}
\def\ee{\end{equation}}
\def\bea{\begin{eqnarray}}
\def\eea{\end{eqnarray}}
\def\lbl{\label}

\begin{document}
\draft           
\title{Slowly rotating voids in cosmology}
\author{Tom\'a\v s Dole\v zel$^{1,2}$, Ji\v r\'\i\ Bi\v c\'ak$^2$ and Nathalie Deruelle$^{1,3,4}$}
\address{$^1$ D\'epartement d'Astrophysique Relativiste et de Cosmologie,\\
Observatoire de Paris--Meudon, UPR 176, CNRS, 92195 Meudon, France \\
$^2$ Institute of Theoretical Physics, Charles University,\\
V Hole\v sovi\v ck\'ach 2, 18000 Prague 8, Czech Republic\\
$^3$ Centre for Mathematical Sciences,DAMTP,\\
University of Cambridge, Wilberforce Road, Cambridge, CB3 0WA, England\\
$^4$ Institut des Hautes Etudes Scientifiques, 91140 Bures-sur-Yvette, France}
\date{\today}

\maketitle

\begin{abstract}
We consider a spacetime consisting of an empty void separated from an almost 
Friedmann-Lema\^\i tre-Robertson-Walker (FLRW)  
dust universe by a spherically symmetric, slowly rotating shell which is comoving with the cosmic dust.
We treat in a unified manner all types of the FLRW universes. 
The metric is expressed in terms
of a constant characterizing the angular momentum of the shell, and parametrized by the comoving radius of the shell.
Treating the rotation as a first order perturbation,
we compute the dragging of inertial frames as well as the apparent motion of distant stars within the void.
Finally, we discuss, in terms of in principle measurable quantities, 'Machian' features of the model.
\end{abstract}


\section*{I Introduction}
The universe at large appears to be remarkably homogeneous and isotropic and its dynamics governed by the gravitational
force created by its material content. In such a FLRW geometry Mach's principle, that is the idea
that local inertial frames are determined by an 'average' motion of the matter in the universe, 
can hardly be put to test because of the too high
symmetry of the model. However the universe is not perfectly homogeneous and isotropic: perturbations are present,
in particular on very large scales, as testified by the  discovery by the COBE satellite of cosmic microwave background
anisotropies as well as the observations of voids with size not negligible with respect to the Hubble radius [1].
Such perturbations, which ought to be treated within a general-relativistic framework, could be used to test, at
least in principle, Mach's ideas - in a similar spirit as they were studied in the case of the FLRW (and
of Lema\^\i tre-Tolman-Bondi) universes in [12].
In particular, if we happened to live in a slowly rotating void, that is an underdense
region separated from the almost FLRW universe outside by a slowly rotating sheet-like
structure or wall,  then the apparent motion of 'distant stars' could be used to
measure the angular velocity of the wall, that is the transverse component of its peculiar velocity, which is
inaccessible to the usual redshift measurements.

As a first approach to this idea of using Mach's principle in observational cosmology we shall here address the
question of how to measure a global rotational velocity field in the universe. To do so we shall study a very simplified
model which however encaptures the physics of the problem: we shall consider an empty void separated from the almost
FLRW (dust) universe outside by a spherically symmetric, slowly rotating shell
comoving with the outer universe. If the assumption of spherical symmetry is reasonable,
the assumption of an empty void, on the other hand, is drastic and will force us to consider very 'heavy' shells which
cannot be treated perturbatively. However, since we are only interested here in the observable effects of the slow
{\it rotation} of the shell, its 'weight' is not so important. Finally the assumption that the shell is comoving with the
cosmic dust means that we consider the void long after its possibly explosive creation, when damping forces have
slowed it down. We shall also ignore all perturbations which are not due to the rotation of the shell as they can
be linearly added.

Such  spacetimes have already been studied in the past. We shall not review here the enormous literature on
{\it non}-rotating voids in cosmology (see e.g. [2] and references therein). Rotating voids in cosmology, 
on the other hand, do not seem to have been much treated.

Indeed, slowly rotating voids in asymptotically {\it flat} spacetimes have mainly been used  to
study  'Machian' effects, how and to what extent they are embodied in general relativity. The classical papers of
Thirring [3], treating a slowly rotating shell of small mass, were generalized by Brill and Cohen [4]  who considered a
slowly rotating shell separating an empty void from the Schwarzschild solution at first order in the rotational
perturbation. Lindblom and Brill [5] further extended their work by treating a freely falling, slowly rotating
shell; in addition, in order to interpret in physical terms the rotational perturbation inside the shell, 
they studied the apparent motion of a 'searchlight' at the centre of the shell as seen from infinity. More
recently, Katz, Lynden-Bell and Bi\v c\'ak [6] solved the opposite problem  (which is more interesting from the
astrophysical point of view) of how fixed stars at infinity appear to rotate to an inertial observer 
at the origin of the
void. An evident drawback of all these Schwarzschild-shell models is that the whole universe is described by the shell
in  an asymptotically flat spacetime so that all rotations are defined with respect to the inertial frames at infinity.
Thus the rotation of all inertial frames within the void is determined, at least in part, by these asymptotic frames that
play the role of something given absolutely, and the shell then cannot be considered as the only source of inertia in the
universe. 

Lewis [8] was the first to discuss the dragging in a truly cosmological context. He considered a slowly
rotating shell immersed  at a given comoving position into a closed FLRW spacetime. He gave the solution for the rotational
perturbation at first order and discussed some Machian features of the model.
Later, Klein [7], analysed more in detail a slowly
rotating shell immersed  at a given comoving position into a FLRW spacetime with open
flat spatial sections $(k=0)$. He discussed the
dragging of inertial frames inside the shell under the condition that the rotational  perturbation is caused by the shell
only. He found that perfect dragging (that is the fact that an inertial observer inside the void measures a 
zero angular velocity for the shell) is reached only as the radius of the shell goes to infinity, 
where the whole matter of the universe
is distributed on the shell. In the next paper [14], Klein also considered some effects caused by slowly rotating shells in $k=0$ 
universes to the second order in the angular velocity.

In section II of the present paper we analyze, in contrast to [7,8], the rotating voids
in {\it all} types of the FLRW universes in a {\it unified} manner, pointing out the differences between the open and closed cases.
We give the spacetime metric as well as all angular velocities in terms of two parameters - of a constant 
characterizing the angular momentum of the shell, and of the comoving radius of the shell. 
Using Bardeen's formalism [10] to treat the rotational perturbation at first order we
calculate the dragging of inertial frames inside the void and determine the conditions for perfect dragging to occur. 
For Einstein-de Sitter universes $(k=0)$ we recover, in an alternative way, the results of Klein [7]. We extend his work to all open models $(k=-1)$ and closed models,
where new features arise, as, for example, the condition that the total angular momentum
of the universe must vanish. We illustrate that perfect dragging cannot be reached in a closed 
universe.

In section III we analyze, in the spirit of Katz, Lynden-Bell and Bi\v c\'ak [6], 
the 
apparent motion of the distant stars as seen by observers inside a void.
As we are aware of, this problem has not been tackled before in a cosmological context.
We thus study the propagation of light in the
spacetimes obtained and describe how the two parameters of the problem, the angular momentum of the
shell and its comoving radius, could in principle be measured.

In the last section IV we
argue that (slow) rotations in the open universes are absolute in the sense
that there is no freedom to choose rotating axes.  
One is thus left with a 'preferred' coordinate system and hence with the possibility to define all rotations 
with respect to this system.
On the other hand, there is no such system in closed universes. 
We finally discuss the Machian features of the model. Aware of the controversies connected
with the formulation and interpretation of Mach's principle [11]
\footnote{For more details on Mach's principle and its various versions see e.g. [11], [16] and ref. therein. A short history
and meaning of Mach's principle is reviewed in [12].},
we express all rotations in terms of, in principle, measurable quantities. The observations of the shell and
the cosmic dust outside the shell, made by inertial observers inside, are in full conformity with Mach's
ideas (for both open and closed universes): the matter
of the models considered is not observed to be globally (slowly) rotating in a preferred direction. In this sense the inertial
frames inside the void are determined by the 'average motion of the matter in the universe' in the spirit of Mach's principle.

Details of calculations of the matching across the shell, of the rotational metric perturbation outside
the shell, and of the angular velocity of a star measured inside the void are given in
Appendices A,B,C.

\section*{II The spacetime metric}

The void is empty so that spacetime 
inside the void can be represented by a portion of Minkowski space. Consider a family of
standard clocks at rest with respect to each other -  they define an inertial frame ${\cal S}_{in}$
with the line element 
\be ds^2|_{in}=-d\bar t^2+d\bar r^2+\bar r^2(d\theta^2+\sin^2\theta\,d\bar\phi^2),\lbl{1}\ee
where $\bar t$ is the time measured by the clocks and $(\bar r,\theta,\bar\phi)$
their spherical coordinates.

Now one can use another time coordinate $t$ such that $d\bar t=\alpha(t) dt$ where the lapse function 
$\alpha(t)$ is to be specified later.
One can also use coordinates which rotate with respect to  $(\bar r,\theta,\bar\phi)$, in
such a way that
$d\bar\phi=d\phi-\tilde\omega_0(t)dt$ where the angular velocity $\tilde\omega_0(t)$ is a function of $t$. In
this new (rotating) frame ${\cal S}_{rot}$ with coordinates  $( t,\bar r,\theta,\phi)$ the line element (\ref{1}) to first
order in $\tilde\omega_0(t)$ reads
\be ds^2|_{in}=-\alpha^2(t)dt^2+d\bar r^2+\bar r^2(d\theta^2+\sin^2\theta\,d\phi^2)-2\bar r^2\,
\tilde\omega_0(t)\,\sin^2\theta\,d\phi\,dt.\lbl{2}\ee

The shell is supposed to be spherically symmetric. Its equation of motion can thus be written as
$\bar r=l(t)$; $l(t)$ is for the moment an arbitrary function of time. The metric induced on the shell by (\ref{2}) is then
\be ds^2|_{in}^\Sigma=-[\alpha^2(t)-\dot{l}^2(t)]dt^2+l^2(t)
(d\theta^2+\sin^2\theta\,d\phi^2-2\tilde\omega_0(t)\,\sin^2\theta\,d\phi\,dt),\lbl{4}\ee
where $\dot{l}\equiv dl/dt$ and $\Sigma$ denotes the shell. 

Outside the void,  the universe is supposed to be almost perfectly homogeneous and isotropic, 
up to a rotational perturbation.
There exists therefore a coordinate system in which the line element reads
\be ds^2|_{out}=-dt^2+a^2(t)\left[{dr^2\over1-kr^2}+r^2(d\theta^2+\sin^2\theta\,d\phi^2)-
2r^2\,\omega(t,r)\sin^2\theta\, d\phi\,dt\right].\lbl{5}\ee
As usual, the curvature index
$k=+1,0,-1$ for respectively spherical, flat or hyperbolic spatial sections, 
$a(t)$ is the scale factor and
function $\omega(t,r)$ is the rotational metric perturbation. 
For the perturbation theory at first order to apply it must be 'small', that is $r\,\omega(t,r) \ll 1$.

We will also suppose that the shell is comoving with the cosmic dust, thus placed at a constant comoving radius $r=l_0$.
The metric induced on such a shell by (\ref{5}) then reads
\be ds^2|_{out}^\Sigma=-dt^2+a^2(t)l_0^2\left[d\theta^2+\sin^2\theta\,d\phi^2-
2\omega_0(t)\,\sin^2\theta\, d\phi\,dt\right],\lbl{9}\ee 
where $\omega_0(t)$ is the value of the rotational perturbation on the shell~: $\omega_0(t)\equiv\omega(t,r=l_0)$.

Let us now complete the matching of the two coordinate grids on the shell.
The induced metrics (\ref{4}) and (\ref{9}) must be continuous [9], [15]. This first gives the 
radial position of the shell $l(t)$ in terms of its comoving coordinate $l_0$ and the scale factor $a(t)$ as
\be l(t)=l_0 a(t).\lbl{10}\ee 
Second it implies that
the angular velocity $\tilde\omega_0(t)$ and the time shift $\alpha(t)$ of the 
coordinate system ${\cal S}_{rot}$ with respect to the system ${\cal S}_{in}$ are
\be \tilde\omega_0(t)=\omega_0(t)\quad,\quad\alpha(t)=\sqrt{1+l_0^2\dot a^2}.\lbl{11}\ee 

Israel's junction conditions [9], [15] 
give the stress-energy tensor of the shell in terms of the jump in its extrinsic curvature (see Appendix A).
If we consider the
shell to be a rotating 2-dimensional perfect fluid, we can write 
\be {\cal T}_{ab}=(\rho_\Sigma+p_\Sigma)w_aw_b+p_\Sigma\gamma_{ab},\quad\hbox{with}\quad
w^a=(1,0,\Omega_\Sigma),\lbl{15}\ee
where $\rho_\Sigma$, $p_\Sigma$ are respectively the shell energy density and surface pressure and
$\Omega_\Sigma\equiv d\phi_\Sigma/dt$ ($\phi_\Sigma(t)$ being the azimuthal coordinate of a fixed point of the shell in the
outside coordinates)
\footnote{At first order in the rotational perturbations $dt=d\tau$, $\tau$ being the proper time on the shell.}
is the  angular velocity of the shell with respect to the outside coordinate grid or, also,
with respect to the rotating frame
${\cal S}_{rot}$ inside the void. 
The indices $(a,b)$ stand for $(t,\theta,\phi)$ and are raised and lowered by the induced metric 
on the shell (\ref{9}).
The matching conditions (App.A, Eq.(\ref{14})) result in
\bea
\rho_\Sigma(t)&=&{1\over4\pi G}{1\over al_0}\left(\sqrt{1+l_0^2\dot a^2}-\sqrt{1-kl_0^2}\right),\nonumber\\
p_\Sigma(t)&=&-{1\over2}\rho_\Sigma-{1\over8\pi G}{l_0\ddot a\over\sqrt{1+l_0^2\dot a^2}},\nonumber\\
\Omega_\Sigma(t)&=&\omega_0(t)-{1\over16\pi G}{\sqrt{1-kl_0^2}\over
a(\rho_\Sigma+p_\Sigma)}\,{\partial\omega\over\partial r}|_0,
\lbl{16}\eea 
where we define $\partial\omega/\partial r|_{0}\equiv \partial\omega/\partial r|_{r=l_0}$.
The energy density and pressure of the shell are determined in terms of $l_0$ and 
the scale factor, and its
angular velocity will be known once the metric perturbation $\omega(t,r)$ is determined. We
note that $\rho_\Sigma(t)$ is always positive and that the equation of state of the shell differs in general from
that of the FLRW universe outside.

A number of global quantities can be associated with the shell. For example we can define its total (proper) mass as
\be m_\Sigma(t)=4\pi a^2l_0^2\rho_\Sigma= {al_0\over G}\left(\sqrt{1+l_0^2\dot
a^2}-\sqrt{1-kl_0^2}\right).\lbl{17}\ee
Note that this mass is {\it not} equal to the (proper) mass $M_\Sigma(t)$ of the universe which
has to be removed in order to create the void. Indeed
\bea
M_\Sigma(t)&=&{a\over2G}\dot a^2l_0^3\quad\hbox{for}\quad k=0\nonumber\\
M_\Sigma(t)&=&{3a\over4G}(1+\dot a^2)\left(\arcsin\l_0-l_0\sqrt{1-l_0^2}\right)\quad\hbox{for}\quad k=1
\lbl{18}\eea 
(and a similar expression for $k=-1$). Only in the the case of voids small compared to the Hubble
and curvature radii ($l_0\dot a\ll 1$, $l_0\ll1$) do $m_\Sigma(t)$ and $M_\Sigma(t)$ coincide. Otherwise, in general,
$M_\Sigma(t)> m_\Sigma(t)$. In any case, unless the void is much larger than the Hubble radius, which is unlikely for causality reasons,
$m_\Sigma(t)$ represents a sizeable fraction of $M_\Sigma(t)$ and therefore is large, which is not surprising since we assumed the void to be totally empty.

More interesting is the angular momentum of the shell
\be j_\Sigma=\int_\Sigma \,{\cal T}^t_{~\phi}\,\sqrt{-\gamma}\,d\phi d\theta\lbl{19},\ee
$\gamma$ being the determinant of the induced metric (\ref{9}), which is conserved due to the axial symmetry
of the problem at hand. Using the decomposition (\ref{15}-\ref{16}) it can be rewritten as
\be j_\Sigma={8\pi\over3}a^4l_0^4(\Omega_\Sigma-\omega_0)(\rho_\Sigma+p_\Sigma)=
-{1\over6G}l_0^4\sqrt{1-kl_0^2}\,a^3{\partial\omega\over\partial r}|_0.\lbl{20}\ee

The source which governs the outside metric (\ref{5}) is the stress-energy tensor of the cosmic  
dust:
\be T_{\mu\nu}=\rho u_\mu u_\nu,\quad\hbox{with}\quad
u^\mu=(1,0,0,\Omega),\lbl{21}\ee
where $g_{\mu\nu}$ is the metric (\ref{5}), 
$\rho$ is the energy density of the dust, and $\Omega(t,r)$ is its (small) angular velocity.

Einstein's equations for the rotational metric perturbation $\omega(t,r)$ are solved in Appendix B. They give
\be \omega(t,r)=-{1\over a^3}(\beta(r)+g(t)),\lbl{25}\ee 
where 
\be
\beta(r)=-{2G\sqrt{1-kr^2}\over r^3}(1+2kr^2)j_\Sigma+6G\int_{l_0}^r{j(r^{\prime})\over
r^{\prime 4}\sqrt{1-kr^{\prime 2}}}\,dr^{\prime}.\lbl{27}
\ee
The radially dependent function $j(r)$,
\be j(r)\equiv\int_{l_0}^r\,T^t_{~\phi}\sqrt{-g}\,d\phi\,d\theta\,dr^{\prime}={8\pi\over
3}a^5\rho\int_{l_0}^r{(\Omega-\omega)\over\sqrt{1-kr^{\prime 2}}}r^{\prime 4}\,dr^{\prime},\lbl{22}\ee 
is the conserved total angular momentum of the dust layer $(l_0,r]$ outside the shell. Hence, the first term on the right hand side
of (\ref{27}), proportional to $j_{\Sigma}$, comes from the rotating shell while the second term is the 
contribution of cosmic matter outside the shell.

As for the function $g(t)$,  it is clear that it can be absorbed in 
the azimuthal coordinate $\phi$ and thus expresses the possibility to choose any global (slowly) rotating axes, i.e. the frame ${\cal S}_{rot}$, but we keep this freedom for later purposes.
Einstein's equations thus yield us the metric perturbation $\omega(t,r)$ in terms of the function of
time $g(t)$ and the radial dependent function $j(r)$. These functions will be determined by boundary conditions. 

\bigskip
$\bullet $ {\it The case of open universes}
\bigskip

In open universes the function $g(t)$ is determined uniquely by the condition
$r\,\omega(t,r)\ll 1$ for all radii $r$ so that the perturbation $\omega(t,r)$
must die away at infinity at least as $r^{-1}$.  When 
$\lim_{r\to\infty}{r\int_{l_0}^{r}j(r^{\prime})dr^{\prime}/(r^{\prime 4}\sqrt{1-kr^{\prime 2}})}$ 
vanishes, as will be the case, 
this amounts, by (\ref{25}) and (\ref{27}), to choosing the
coordinate system in such a way that
\be g(t)=0\quad\hbox{for}\quad k=0;\qquad g(t)=-4Gj_\Sigma\quad\hbox{for}\quad k=-1.\lbl{28}\ee
The metric (\ref{2}) inside the void is then completely known: 
$\alpha(t)$ and $\tilde\omega_0(t)$ are given by (\ref{11}) and 
\bea
\omega_0(t)&=&{2Gj_\Sigma\over a^3l_0^3}\quad\hbox{for}\quad k=0\nonumber\\
\omega_0(t)&=&{2Gj_\Sigma\over a^3l_0^3}\left[\sqrt{1+l_0^2}(1-2l_0^2)+2l_0^3\right]\quad\hbox{for}\quad k=-1.
\lbl{29}\eea

Let us finally turn to the metric outside the shell which still depends on the (so far unknown) function $j(r)$. 
The conservation laws for the dust universe imply $\rho\propto a^{-3}$ so that from (\ref{22}) we deduce that the  
combination $a^2(\Omega-\omega)$ does not depend on time.
We thus define a function $f(r)\equiv a^2(\Omega-\omega)$ which describes the perturbation of the stress-energy tensor 
of cosmic dust outside the shell, $T^t_{~\phi}=\rho(t) r^2\sin^2\theta f(r)$, or, equivalently, its 
angular momentum distribution.
In general it is not determined uniquely just by requiring $\omega\rightarrow 0$ at infinity 
and can in principle be fixed by initial conditions or deduced from observations. 
However, if we demand, in accord with Klein [7], that the only source for the rotational perturbation $\omega$ at any $r$
is the shell itself (brought in some way into a slow rotation), we eliminate the contribution of the 
cosmic dust in (\ref{27}) by choosing $f(r)=0$. Then
\be \Omega(t,r)=\omega(t,r)\quad\hbox{i.e.}\quad j(r)=0.\lbl{31}\ee 

With the condition (\ref{31}) the metric outside the shell is known in terms of
the single parameter $j_\Sigma$:
\bea
\omega(t,r)&=&{2Gj_\Sigma\over a^3r^3}\quad\hbox{for}\quad k=0\nonumber\\
\omega(t,r)&=&{2Gj_\Sigma\over a^3r^3}\left[\sqrt{1+r^2}(1-2r^2)+2r^3\right]\quad\hbox{for}\quad k=-1.
\lbl{32}\eea 

\bigskip
$\bullet $ {\it The case of closed universes}
\bigskip

In this case the condition $r\,\omega(t,r)\ll 1$ holds for any small $g(t)$ 
(with $\beta(r)$ small - cf. (\ref{25})) and we are left with
a freedom to choose (slowly) rotating axes. We can thus choose the gauge simply as
\be g(t)=0.\lbl{33}\ee
The metric inside the void is then known. In particular
\be \omega_0(t)={2Gj_\Sigma\over a^3l_0^3}\sqrt{1-l_0^2}(1+2l_0^2).\lbl{34}\ee

In order now to determine the metric outside the void  we first take into account the fact that the total
angular momentum of the {\it closed} universe
must vanish [7], [12]. The choice $f(r)\equiv a^2(\Omega-\omega)=0$, used in the open cases
is not compatible with this constraint. This means that the integral in (\ref{27}) cannot be eliminated, 
i.e. the matter outside
the shell {\it does} necessarily influence the rotational perturbation: the cosmic dust has to have some intrinsic 
angular momentum to counteract the rotation of the shell.
Nevertheless, even the zero total angular momentum condition does not 
determine the function $f(r)$ uniquely.    
We shall, for definiteness, treat the example $f=constant$, where the constant is fixed by  
$j(\pi)=-j_\Sigma= (8\pi f/3)(\rho a^3)\int_{\chi_0}^\pi \sin^4\chi d\chi$ where
we put as usual $\sin\chi=r$ ($\sin\chi_0 = l_0$). 
Then $a^2(\Omega-\omega)\propto j_{\Sigma}$ and the metric outside the shell is known in terms of the angular momentum of the shell $j_\Sigma$.
We obtain
$$ f=-{j_\Sigma}\left[{8\pi\over 3}(\rho a^3)(\xi(\pi)-\xi(\chi_0))\right]^{-1},$$
where
\be\xi(\chi)={1\over32}(12\chi-8\sin2\chi+\sin4\chi),\lbl{35}\ee 
and the analogue of condition (\ref{31}) reads 
$$\Omega(t,r)=\omega(t,r)-j_\Sigma\left[{8\pi\over 3}(\rho a^5)(\xi(\pi)-\xi(\chi_0))\right]^{-1},$$
i.e.
\be j(\chi)=-j_\Sigma{\xi(\chi)-\xi(\chi_0)\over\xi(\pi)-\xi(\chi_0)}.\lbl{36}\ee 
Equation (\ref{27a}) can now be integrated to obtain the rotational perturbation outside the shell in the form
\be \omega(t,\chi)= {3\over8} {{Gj_\Sigma}\over{\xi(\pi)-\xi(\chi_0)}} {1\over a^3}
\left\{{1\over \sin^3\chi}\left[ 2\sin\chi+(\pi-\chi)\left(3\cos\chi-\cos3\chi\right)\right]
- {\cal C}(\chi_0)\right\},\lbl{37}\ee
where ${\cal C}(\chi_0)$ is given by
\be {\cal C}(\chi_0)= {1\over\sin^3\chi_0}\left\{\sin2\chi_0({1\over2}+{1\over3}\sin^2\chi_0)(\cos3\chi_0-
3\cos\chi_0)+2\sin\chi_0\right\},\lbl{38}\ee
the comoving radius of the shell, $\chi_0 \in (0,\pi)$, being a parameter of the solution. The
value of the rotational perturbation at the shell is obtained by the limit $\chi \rightarrow \chi_0$ in (\ref{37})
\be \omega_0(t)=-{Gj_{\Sigma}\over a^3\sin^3\chi_0}[\cos3\chi_0-3\cos\chi_0],\lbl{39}\ee 
which is identical to (\ref{34}).
The function $a^3\omega_0$ decreases monotonously as the function of the comoving
radius of the shell $\chi_0$; it vanishes at $\chi_0={\pi\over 2}$. Fig.\ref{fig1} shows the rotational
perturbation $a^3\omega$ as a function of the radial coordinate $\chi$ for several
positions $\chi_0$ of the shell. For $\chi_0 < {\pi \over 2}$ the rotational perturbation vanishes at some point, say
$\tilde\chi \in [\chi_0,\pi]$. For $\chi_0 > {\pi \over 2}$ the rotational perturbation is negative in the
whole spacetime outside the shell. However, we can always choose the function $g(t)$
in such a way that $a^3\omega$ vanishes at some given $\tilde\chi$ whatever is the
value of $\chi_0$ - see section IV for a discussion of this point. 

\begin{figure}[h]
\centerline{\psfig{figure=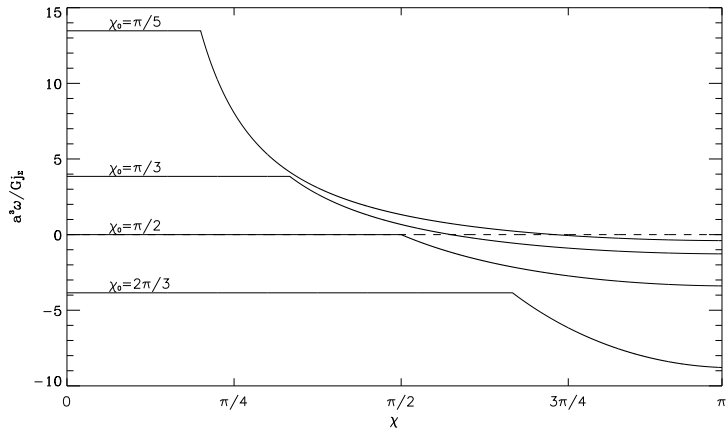,width=4in}}
\caption{ \it Rotational perturbation $a^3\omega/Gj_{\Sigma}$ as a function of radial coordinate $\chi$ 
for several comoving radii of the shell $\chi_0$ $(k=+1)$. Inside the void, $\chi<\chi_0$, the perturbation is constant, 
given by (\ref{39}). The curves are plotted in the gauge corresponding to $g(t)=0$; any non-zero $g(t)=constant$ 
would shift the picture along the y-axis.}
\lbl{fig1}
\end{figure}

\section*{III Angular velocity of the shell and apparent motion of  distant stars}

\subsection*{a The proper angular velocity of the shell}

Let us now turn to the rotation of the shell. 
In (\ref{15}) we defined the shell angular velocity with 
respect to the outside coordinate grid: $\Omega_\Sigma\equiv d\phi_\Sigma/dt$. 
We can also define its proper angular velocity,
$\bar\Omega_\tau\equiv d\bar\phi_{\Sigma}/ d\tau$, $\bar\phi_\Sigma(t)$ being the trajectory of a point of the
shell in the inertial frame ${\cal S}_{in}$ and $\tau$ being its proper time. As the shell is slowly rotating,
we see at first order, i.e. for $(l_0 a)\bar\Omega_{\tau} \ll 1$ and $d\tau=dt$, that
\be \bar\Omega_{\tau}=\Omega_{\Sigma}-\omega_0. \lbl{A2}\ee
Putting together equations (\ref{16}), (\ref{25}) and (\ref{27a}), and recalling that
$\lim_{r\to l_{0_+}}{j(r)}=0$, we find for all $k=-1,0,1$:
\be \bar\Omega_{\tau}={3Gj_\Sigma\over a^3l_0^3}{\sqrt{1+l_0^2\dot a^2}\over 
1-\sqrt{1+l_0^2\dot a^2}\sqrt{1-kl_0^2}-l_0^2a^2(\dot a/a)\dot{~}}\,\,.\lbl{A3}\ee
Note that in order to keep $j_\Sigma$ constant, the agular velocity of the shell in general depends on time through the
scale factor $a(t)$. 

\subsection*{b The apparent motion of distant stars}

A {\it fixed star}, as defined in [6] in the case of the Schwarzschild shell, denotes a source of light placed at infinity
at rest with respect to the asymptotic inertial frame on the centre of the shell. We shall not in general consider such fixed 
stars as they are of little use in a cosmological context for the following reasons: 1. the radius of the visible 
universe is limited by the horizon; 2. there is no infinity in closed cosmological models; 3.  'stars' comove with 
the cosmic matter and in general they are not at rest with respect to the $(r,\theta, \phi)$ coordinate grid
because of the presence of the (time-dependent) rotational perturbation. 
We shall thus talk about the apparent motion of {\it distant stars} rather than fixed stars, the 4-velocity of these 
stars being given by the 4-velocity of the cosmic dust, i.e. by (\ref{21}). 

The {\it angular velocity of a distant star as measured by an observer in the inertial frame} ${\cal S}_{in}$ 
in the void is given by
\be \bar\Omega_{star}^{m}\equiv{\delta\bar\phi_{rec}\over\delta\bar t_{rec}}\lbl{A8},\ee
where $\delta\bar t_{rec}\equiv\bar t^2_{rec}-\bar t^1_{rec}$ is observer's proper time interval between the arrival
times of two photons emitted by the star and $\delta\bar\phi_{rec}\equiv\bar\phi_{star}(\bar
t^2_{rec})-\bar\phi_{star}(\bar t^1_{rec})$ is the angular separation travelled by the star in
the interval $\delta\bar t_{rec}$, as measured in ${\cal S}_{in}$.
The results of Appendix C show that this angular velocity 
of a star comoving with the cosmic dust at a given radius $\chi_*$ (where $r=\chi,\sinh\chi,\sin\chi$ for $k=0,-1,+1$, respectively),
emitting photons radially inwards in the equatorial plane,
\footnote{Putting the observer off the centre complicates the calculation by introducing a radial velocity of the star
in which we are not interested here.} is given by the expression
\be \bar\Omega_{star}^{m}={1\over[a(\alpha+l_0\dot a)]_0}\left([a\Omega]_*-[a\omega]_*+
\int_{\chi_0}^{\chi_*}{d(a^3\omega)\over d\chi}{1\over a^2}d\chi\right).\lbl{A18}\ee 
The subscripts '*' and '0' mean that the quantities are evaluated at the time of emission and at the time when the first photon
reaches the shell, respectively. We recall that the comoving radius of the shell is $l_0=\chi_0,\sinh\chi_0$ or $\sin\chi_0$.

\bigskip
$\bullet $ {\it The case of open universes}
\bigskip

The scale factor of a $k=0$ universe is given in terms of the conformal time $\eta$ and a constant
$a_M$ as 
$a(\eta)={1\over 4}a_M\eta^2$.
With the condition (\ref{31}) that the rotational perturbation arises only due to the rotation of the shell,
$\Omega(\eta_*,\chi_*)=\omega(\eta_*,\chi_*)$, 
we obtain from Eq.(\ref{A18}) staightforwardly
\be \bar\Omega_{star}^m=-{384Gj_{\Sigma} \over a_M^3\eta_0^2}{1\over \sqrt{1+4\chi^2_0/\eta^2_0}+2\chi_0/\eta_0}
\int_{\chi_0}^{\chi_*}{d\chi\over \chi^4 [\chi_0+\eta_0-\chi]^4},\lbl{A19}\ee
where $\eta_0=\chi_*+\eta_*-\chi_0$. 
Since $\bar\Omega_{star}^m$ can never be positive, the distant stars are seen rotating {\it backwards}. Note
that the stars placed just behind the shell,
$\chi_*\to\chi_0$, do not rotate with respect to the inertial observers inside the void.

Interesting from the observational point of view is today's dependence of 
$\bar\Omega_{star}^m/j_{\Sigma}$ on the cosmological redshift $z(\chi_*)$ of the star.
$\eta_0$ is given as a function of the conformal time $\eta_{rec}$ elapsed by the time
$\bar{t}_{rec}$ when the first ray reaches the observer at the centre of the void. Between $\bar{t}_0$
and $\bar{t}_{rec}$, when the light travels inside the void, the conformal time changes from $\eta_0$ to
$\eta_{rec}$ so that
$$a(\eta_0)\chi_0={1\over 4}a_M\eta_0^2\chi_0 = \int_{\bar{t}_0}^{\bar{t}_{rec}}d\bar{t}
=\int_{\eta_0}^{\eta_{rec}}a(\eta)\alpha(\eta)d\eta
={1\over 4}a_M\int^{\eta_{rec}}_{\eta_0}\eta^2\sqrt{1+4{\chi_0^2 \over \eta^2}}d\eta,$$     
resulting in 
\be (\eta_{rec}^2+4\chi_0^2)^{3/2}=3\chi_0\eta_0^2+(\eta_0^2+4\chi_0^2)^{3/2},\lbl{A20}\ee  
and $\eta_{rec}$ is given as a function of today's estimated age
of the universe, $t_{today}=1.1\times 10^{10}$~years, as
$\eta_{rec}=\left(12t_{today}/ a_M\right)^{1/3}.$
Equation (\ref{A20}) gives the expression for $\eta_0$ explicitly as
$$\eta_0^2=\left\{\left[X+Y\right]^{1/3}+
                  \left[X-Y\right]^{1/3}-\chi_0\right\}^2-4\chi_0^2,$$
in which
$$X={1\over 2}\left(\eta_{rec}^2+4\chi_0^2\right)^{3/2}+5\chi_0^3\quad\hbox{and}\quad
  Y=\sqrt{{1\over 4}\left[(\eta_{rec}^2+4\chi_0^2)^{3/2}+10\chi_0^3\right]^2-\chi_0^6}.$$ 
The equality (\ref{A19}) can now be evaluated as a dependence  $\bar\Omega_{star}^m/j_{\Sigma}\,(z,t_{today})$,
$z$ being the measured redshift,
$$z(\chi_*)\equiv {a(\eta_0)\over a(\eta_0+\chi_0-\chi_*)}-1
            ={\eta_0^2\over \left[\eta_0+\chi_0-\chi_*\right]^2}-1.$$ 
Several curves illustrating today's angular velocity of a star parametrized by the comoving 
radius of the shell, as a function of $z$, 
are plotted in Fig.\ref{fig2}. As the measured redshift depends on $a(\eta_0)$, when
the light ray leaves outer universe, and not on $a(\eta_{rec})$ at the time of observation, 
the cosmological redshift does not bring any information about the radius of the void.

Eq.(\ref{A19}) also shows that in a nonstationary universe  
the distant stars do not rotate uniformly: for $\chi_0$, $\chi_*$ fixed,
$\bar\Omega_{star}^m$ depends on the time of observation through $\eta_0$ and (\ref{A20}).
$\bar\Omega_{star}^m$ decreases monotonously down to zero as $\eta_0\rightarrow\infty$
(then $a\rightarrow\infty$ so that the rotational perturbation vanishes).   

The $k=-1$ models give qualitatively the same results.

\begin{figure}[ht]
\begin{center}
\leavevmode
\epsfxsize=3.1in \epsfbox{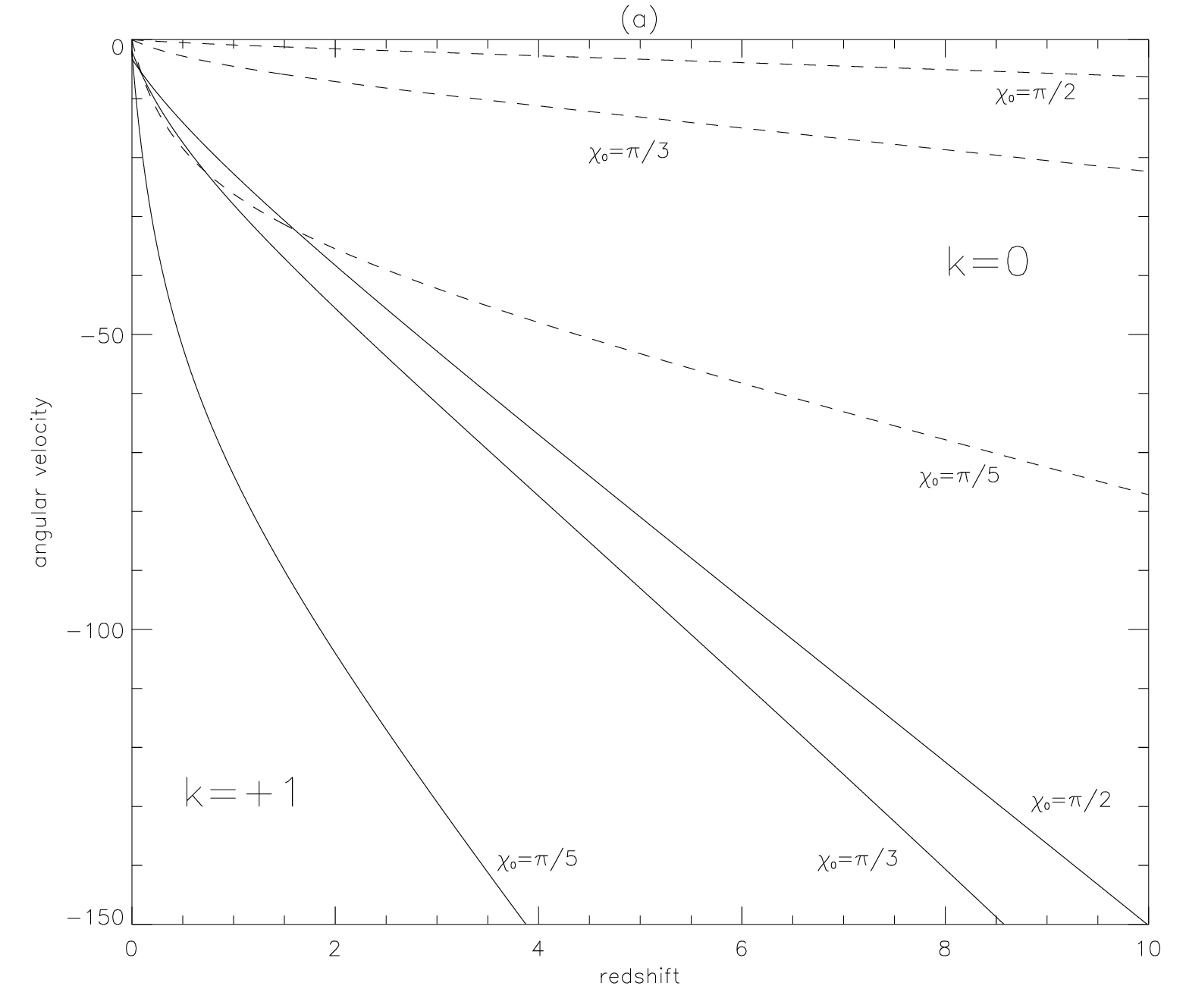} 
\epsfxsize=3.1in \epsfbox{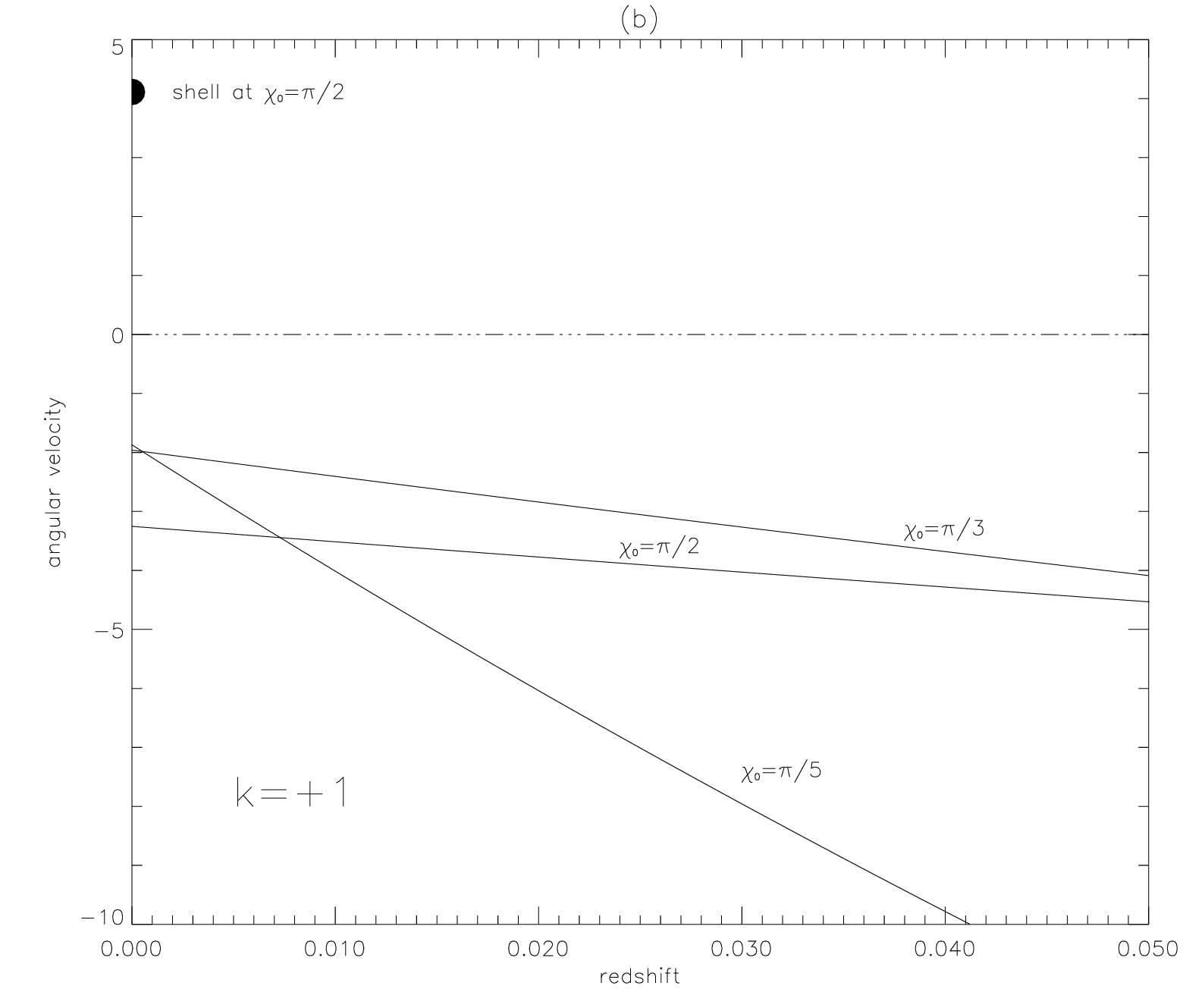}
 \end{center}
\caption{\lbl{fig2} \it (a) Observed angular velocity of the cosmic matter
$\bar\Omega^m_{star} a^3_M/Gj_{\Sigma}$ as a function
of the redshift, plotted for both open (dashed lines) and closed (solid lines) universes, and
corresponding to today's time of observation. 
(b) The near vicinity of the shell for $k=+1$ is shown. While for $k=0$ all the curves start at the origin,
the starting point of each curve is negative when $k=+1$, and depends on the radius of the shell.
The 'discontinuity' between the angular velocity of the shell and those of the stars just behind the shell,   
$\bar\Omega_{star}^m \not\rightarrow \bar\Omega_{\Sigma}^m$ 
in the limit $\chi_*\rightarrow\chi_0$, is shown for $\chi_0={\pi\over 2}$.}
\end{figure}

\bigskip 
{\it Remark}: 
One can check in Eq.(\ref{A20}) that $\chi_0+\eta_0>\eta_{rec}$ when $\chi_0>0$, and $\eta_0=\eta_{rec}$ for $\chi_0=0$. 
Since $\chi_0+\eta_0$ corresponds to the coordinate value
of the particle horizon of an observer at the centre of the void at the time $t_{today}$, while $\eta_{rec}$
would correspond to today's particle horizon of a pure FLRW observer, the presence of a void
increases the particle horizon of the observers inside.       

\bigskip
$\bullet $ {\it The case of closed universes}
\bigskip

In the closed case also the cosmic dust contributes to the rotational perturbation: 
$\Omega(\eta_*,\chi_*)-\omega(\eta_*,\chi_*)\neq 0$.
Equations (\ref{27a}) and (\ref{36}) imply 
\bea a(\eta_*)[\Omega(\eta_*,\chi_*)-\omega(\eta_*,\chi_*)]&=&
-{j_{\Sigma}G\over[\xi(\pi)-\xi(\chi_0)]}{1\over a_M a(\eta_*)},\nonumber\\
{d [a^3\omega](\chi)\over d\chi}&=&-{6Gj_{\Sigma}\over \sin^4\chi}{\xi(\pi)-\xi(\chi)\over \xi(\pi)-\xi(\chi_0)},\nonumber\eea
so that Eq.(\ref{A18}) becomes
\be \bar\Omega_{star}^m=
-{4Gj_{\Sigma}\over [\xi(\pi)-\xi(\chi_0)]a^3_M}
\left[{1\over \alpha(\eta_0)(1-\cos\eta_0)+\sin\eta_0\sin\chi_0}\right]
\left[{1\over 1-\cos[\chi_*-\chi_0-\eta_0]}+{\cal I}\right],
\lbl{A21}\ee 
where
$${\cal I}\equiv 12\,\int_{\chi_0}^{\chi_*}{\xi(\pi)-\xi(\chi)\over
\sin^4\chi(1-\cos[\chi-\chi_0-\eta_0])^2}d\chi\quad\hbox{and}\quad
\alpha(\eta_0)=\sqrt{1+{\sin^2\chi_0\sin^2\eta_0\over(1-\cos\eta_0)^2}}.$$

$\bar\Omega_{star}^m$ in the above expression is negative for all $\chi_*$, $\chi_0$, $\eta_0$. 
As in the case of open universes, the distant stars
are seen rotating {\it backwards}. In contrast with the open case,
the cosmic dust induces the first term at the right hand
side of (\ref{A21}), which does not vanish, unlike the integral ${\cal I}$, in the limit $\chi_*\to\chi_0$. 
Hence, if the dust outside the shell contributes to the rotational perturbation,  
the stars just behind the shell in general rotate 
with respect to the inertial observers inside the void (see Fig.\ref{fig2}).

The dependence $\eta_0$ on $\eta_{rec}$ (obtained similarly to the case $k=0$), is given by
$$a(\eta_0)l_0={a_M\over2}(1-\cos\eta_0)\sin\chi_0={a_M\over 2}\int^{\eta_{rec}}_{\eta_0}(1-\cos\eta)
                 \alpha(\eta)\,d\eta,$$
which can be solved numerically (only when $\chi_0=\pi/2$ one can obtain a simple analytic formula
for $\eta_0(\eta_{rec})$).
Inserting  $\eta_0(\eta_{rec})$ into (\ref{A21}), together with
$$t_{today} = {a_M\over 2}\left(\eta_{rec}-\sin\eta_{rec}\right),\quad 
z={1-\cos\eta_0\over 1-\cos[\eta_0+\chi_0-\chi_*]}-1,$$
we get today's dependence $\bar\Omega_{star}^m/j_{\Sigma}$ on the redshift;      
Fig.\ref{fig2} shows several typical curves parametrized by $\chi_0$.
 
\subsection*{c The apparent motion of the shell}

Equation (\ref{A18}) allows to determine the angular velocity of the shell as measured by the inertial observer 
at the centre of the void. If one replaces $\Omega_*$ by $\Omega_{\Sigma}$,
and with $\eta_*=\eta_0$, one gets the apparent angular velocity of the shell as
\be \bar\Omega_{\Sigma}^{m}={1\over[\alpha+l_0\dot a]_0}[\Omega_{\Sigma}-\omega_0]_0={1\over[\alpha+l_0\dot a]_0}
[\bar\Omega_{\tau}]_0, \lbl{A22}\ee
$\bar\Omega_{\tau}$ is the proper angular velocity defined in subsection IIIa.

\bigskip
$\bullet $ {\it The case of open universes}
\bigskip

Combining the above expression with the expression for the proper angular velocity of the shell (\ref{A3}) we find 
\be \bar\Omega_{\Sigma}^m={192Gj_{\Sigma}\over a^3_M\chi^3_0\eta^6_0}{\sqrt{1+4\chi_0^2/\eta_0^2}\over
[2\chi_0/\eta_0+\sqrt{1+4\chi_0^2/\eta_0^2}][1+6\chi^2_0/\eta^2_0-\sqrt{1+4\chi_0^2/\eta_0^2}]}.\lbl{A23}\ee 
This expression can never be negative. Thus, for any given finite $\chi_0$ the shell is seen rotating
{\it forwards}. It decreases as  
$\chi_0$ increases and vanishes as $\chi_0\rightarrow\infty$. A similar behaviour is obtained for $\chi_0$
fixed in the limit $\eta_0\rightarrow\infty$ (Fig.\ref{fig4}). 

The case of $k=-1$ open universes yields qualitatively the same results.

\bigskip
$\bullet $ {\it The case of closed universes}
\bigskip

Repeating the same steps as in the previous case we now get 
\be \bar\Omega_{\Sigma}^m={24Gj_{\Sigma} \over a_M^3 \sin^3\chi_0[1-\cos\eta_0]^3}
{\alpha(\eta_0)\over\left[{\sin\chi_0\sin\eta_0\over 1-\cos\eta_0}+\alpha(\eta_0)\right]}
{1\over \left[\alpha^2(\eta_0)-\cos\chi_0\alpha(\eta_0)+{\sin^2\chi_0\over 1-\cos\eta_0}\right]}.
\lbl{A24}\ee 
It is easy to see that, just as in the case of open universes, $\bar\Omega_{\Sigma}^m$ can never be
negative. The shell is thus always seen rotating {\it forwards}.
However, in contrast to the open case, $\bar\Omega_{\Sigma}^m$, as a function of $\eta_0$,
shows a minimum which never reaches zero. One gets a similar behaviour for $\eta_0$ fixed and varying 
$\chi_0$ (Fig.\ref{fig4}). We will return to this point in the next section.

\bigskip
The discontinuity in the angular velocities at the shell, $\bar\Omega_{star}^m \not\rightarrow \bar\Omega_{\Sigma}^m$ 
in the limit $\chi_*\rightarrow\chi_0$,
shown in Fig.\ref{fig2}, allows one to determine
observationally, at least in principle, the two parameters of the problem: $j_{\Sigma}$ and
$l_0$. 
The light emitted from sources placed on the shell, and just behind it, propagates in Minkowski spacetime and
is observed with zero cosmological redshift. The source rotating forwards must be on the shell; the measurement
of $\bar\Omega_{\Sigma}^m$ in Eq.(\ref{A23}) or (\ref{A24}) gives  $j_{\Sigma}$ as a function of $\chi_0$.
In the case $k=1$, an additional measurement of a counter-rotating $z=0$ source 
$\bar\Omega_{star}^m(\chi_*\rightarrow\chi_0)$ determines, due to (\ref{A21}), both $j_{\Sigma}$ and $\chi_0$. 
For $k=0$ Eq.(\ref{A19}) implies that
$\bar\Omega_{star}^m(\chi_*\rightarrow\chi_0) \rightarrow 0$, and the observer inside the void must therefore
measure a source at a known redshift instead of a source just behind the shell.
$\bar\Omega_{star}^m(z)$, where the redshift corresponds to the distance between $\chi_0$ and $\chi_*$,   
determines, together with  $\bar\Omega_{\Sigma}^m$, both $j_{\Sigma}$ and $\chi_0$.

One can alternatively measure the angular velocities of distant stars at known redshifts and 
read off the parameters $j_{\Sigma}$
and $\chi_0$ from the curves at Fig.\ref{fig2}. The angular velocity of the shell then follows from
(\ref{A23}) or (\ref{A24}).
For given values of  $j_{\Sigma}$ and $\chi_0$, the angular velocities of distant stars at known redshifts depend
on the type of the FLRW model so that their measurement can also in principle determine the curvature index $k$.

It is clear that above results depend quantitatively on our choice of the function $f(r)$ expressing the initial condition
on the motion of the cosmic matter outside the shell - 
another choice in (\ref{31}) and (\ref{35}) would have led to a different
redshift dependence of $\bar\Omega^m_{star}$. The actual measurement of the redshift dependence of  $\bar\Omega^m_{star}$
would then determine $f(r)$.
Note however that the angular velocity of the shell, $\bar\Omega_{\Sigma}^m$, does not depend on the choice
of $f(r)$, see Eq.(\ref{A22}). 
We note also that since $a^2(\Omega_{\Sigma}-\omega_0)$ depends on time, we cannot
require $\lim_{\chi\to\chi_0}{f}=a^2(\Omega_{\Sigma}-\omega_0)$.   
Hence, a discontinuity in the angular velocities $\bar\Omega_{\Sigma}^m$, $\bar\Omega_{star}^m$
at the shell occurs for any choice of $f(r)$. 

The determination of the void radius, allowed only
thanks to the presence of the rotational perturbation, is crucial for the observer inside the void. Only if 
$\chi_0$ is known, one can properly adjust the redshift method to measure distances - this would not be
possible inside a void in a non-perturbed FLRW universe. 

\section*{IV Dragging inside the void and Mach's principle}

The inertial frame ${\cal S}_{in}$ attached to the centre of the void, as well as other inertial frames in the void
all rotate with the same angular velocity with 
respect to the rotating frame
${\cal S}_{rot}$ or, equivalently, with respect to the observers
inside the shell fixed at constant $\bar r,\theta,\phi$. 
The two coordinate grids are related by 
$d\bar\phi=d\phi-\omega_0(t)dt$.
From $\bar\phi=constant$ it follows that
\be {d\phi\over dt}=\omega_0(t),\lbl{A1}\ee 
where the explicit expression for $\omega_0(t)$ is (\ref{29}) or (\ref{34}). 
The rotation of the spatial axes of the frame ${\cal S}_{in}$ 
with respect to those of ${\cal S}_{rot}$ demonstrates {\it the dragging of inertial frames} inside the void. 
It will be useful to define the {\it preferred} observers outside the shell as those which are at rest in the outside 
coordinate grid in which the line element is given by (\ref{5}), i.e. at constant $r,\theta,\phi$.  
The preferred observers are not in general freely falling; 
in expanding and collapsing stages when $\dot a/a\neq 0$ they fall freely at the radii
where the rotational perturbation vanishes.   
The choice of $g(t)$ in (\ref{28}) and (\ref{33}) implies that 
the preferred observers fall freely at infinity in open cases, and at a certain 
$\tilde\chi$ in closed cases 
(see discussion below Eq.(\ref{39})).

In the open universes the rotating axes were fixed uniquely by the condition $r\,\omega(t,r)\ll 1$. In this frame
physically realizable preferred observers cover the whole outer universe and can be defined as 
{\it globally non-rotating}. The frame dragging inside a void in the open universes can thus be 
regarded as {\it absolute} with respect to these observers.  

In a closed universe, on the other hand, the value of the rotational perturbation at the shell, $\omega_0(t)$, 
is a gauge dependent quantity. The function
$g(t)$ can be used to choose any arbitrary path $\tilde\chi(t)$ as a locus of freely falling preferred observers, 
or to choose any value for $\omega_0(t)$ (cf. Fig.\ref{fig1}). 
In particular, by choosing $g(t)=constant$ $\equiv -\beta(l_0)$, we obtain $\omega_0(t)$ vanishing at all times.
In this case the preferred observers fall freely at the shell and the 
spatial axes of the two frames ${\cal S}_{in}$ and ${\cal S}_{rot}$ coincide. Hence, the dragging of inertial frames
inside a void in closed universes is {\it relative}, i.e. depending on the choice of gauge. 

Now, since we have expressed all rotations in terms of mesurable quantities $\bar\Omega_{star}^m$ and $\bar\Omega_{\Sigma}^m$
(their gauge invariance being evident since $g(t)$ cancels out in both
($\Omega-\omega$) and ($\Omega_{\Sigma}-\omega_0)$) 
we shall confront the observation of the angular velocity distribution of matter in the 
universe, made by inertial observers inside the void, directly with Mach's ideas. By
{\it perfect dragging} we shall mean that:  
an inertial observer inside the void  is {\it dragged perfectly by the shell} 
if the angular velocity of the shell $\bar\Omega_{\Sigma}^m$, measured in the frame
${\cal S}_{in}$, vanishes.   

\begin{figure}[ht]
\begin{center}
\leavevmode
\epsfxsize=3in \epsfbox{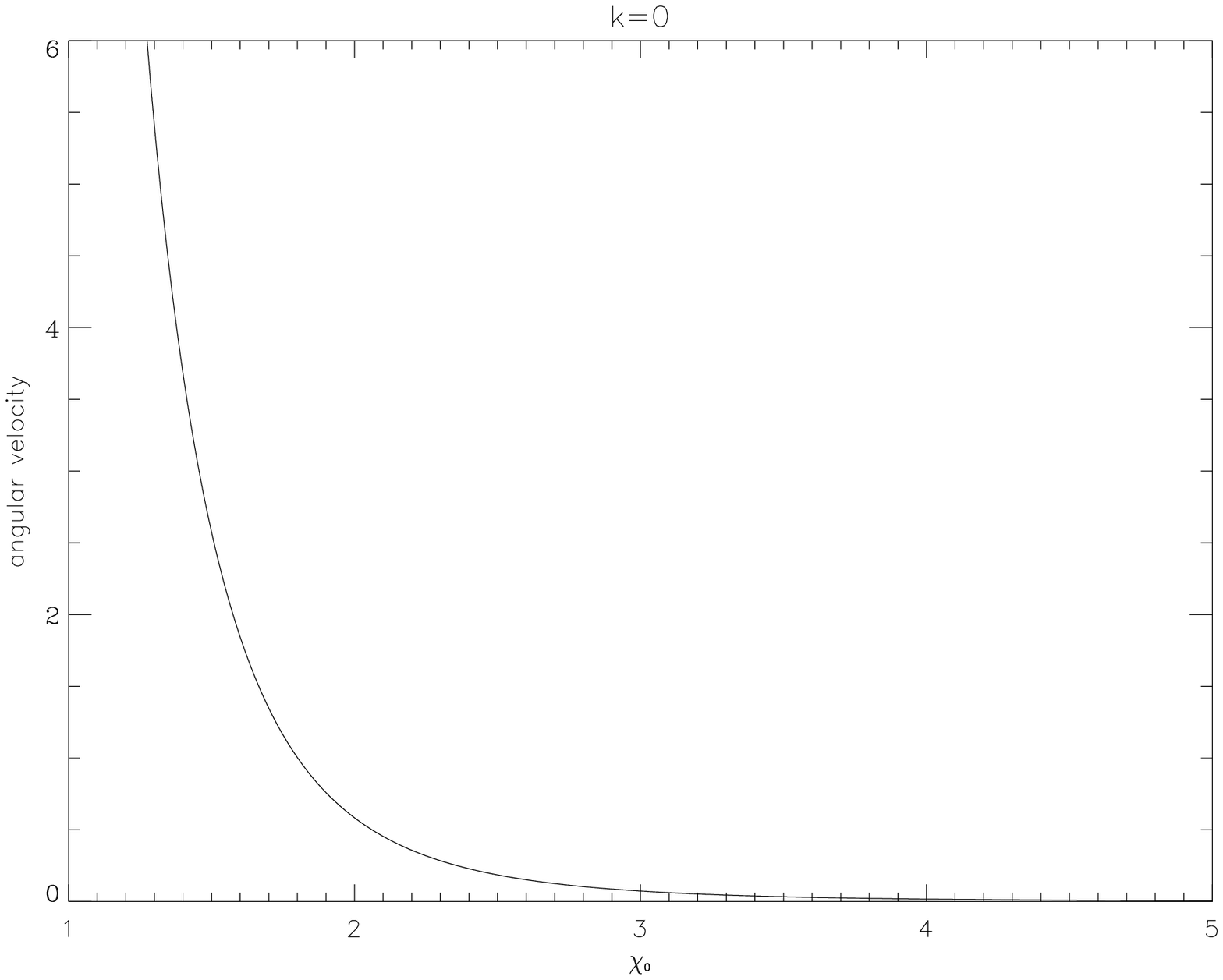} 
\epsfxsize=3in \epsfbox{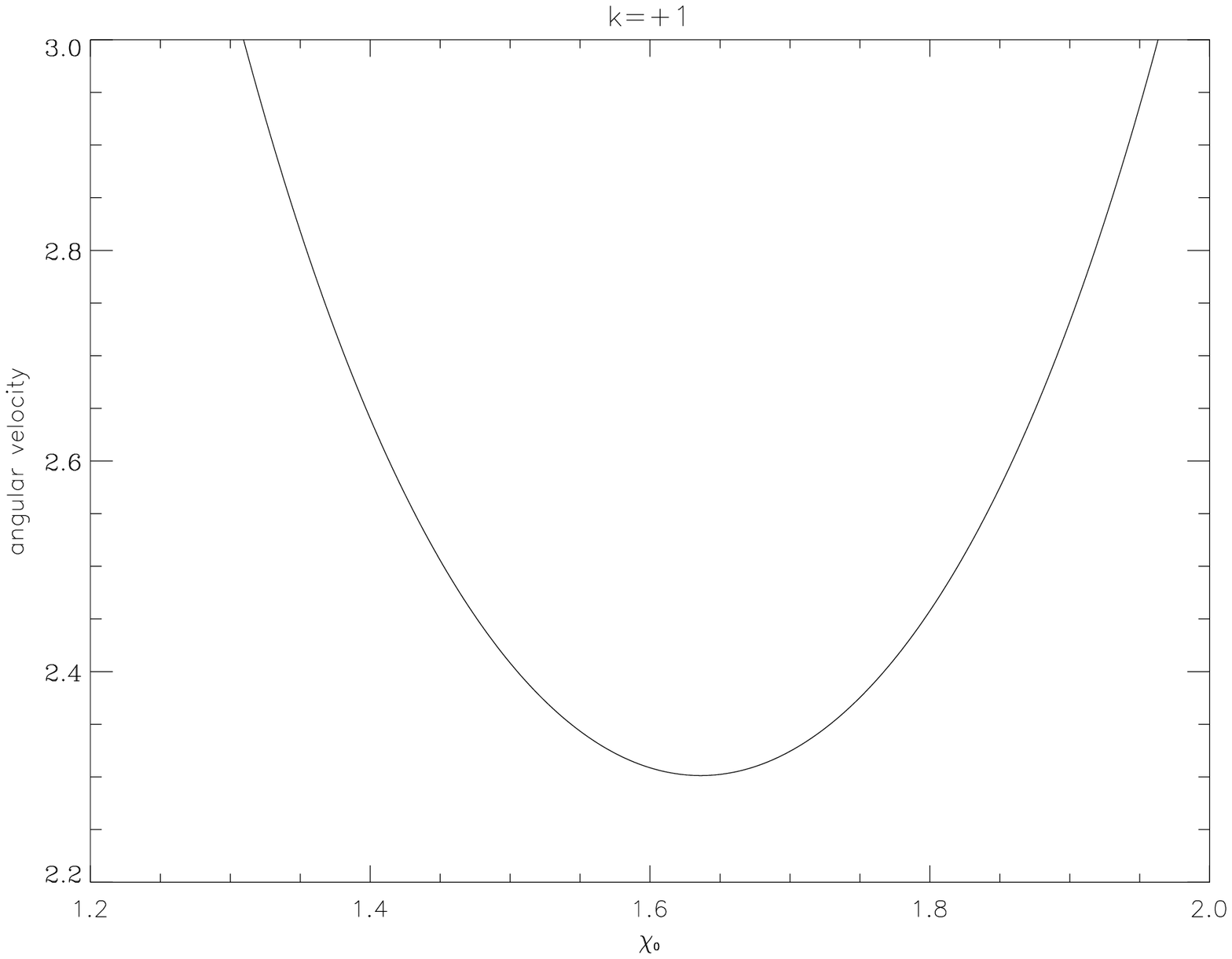} 
\end{center}
\caption{\lbl{fig4} \it Observed angular velocity of the shell, $\bar\Omega^m_{\Sigma} a^3_M/Gj_{\Sigma}$, as a
function the comoving radius $\chi_0$; $\eta_0=1$. When $k=0$, $\bar\Omega_{\Sigma}^m$ vanishes as 
$\chi_0\rightarrow\infty$; 
for $k=+1$, $\bar\Omega_{\Sigma}^m$ shows a positive minimum. A similar behaviour is obtained for $\chi_0$ fixed and 
$\eta_0$ varying.}     
\end{figure}

\bigskip
$\bullet $ {\it The case of open universes}
\bigskip

In the case $k=0$ (for $k=-1$ we get similar results) $\bar\Omega_{\Sigma}^m$   
decreases as the radius of the shell $\chi_0$ or the expansion factor $a$ increase, and
dies away as 1) $\chi_0 \rightarrow \infty$ for $a$ fixed  (Fig.\ref{fig4}) or 
2) $\eta \rightarrow \infty$ for $\chi_0$ fixed.
This behaviour is in agreement with Klein's result [7] and can be interpreted from a Machian point of view
as follows:

1) When the radius of the shell is getting large, almost all matter of the universe can be regarded
as redistributed on the shell. In this case, the angular velocity of the shell fully determines the angular 
velocity distribution of the matter in the universe and the frame dragging becomes perfect. This is in full
conformity with Mach's ideas since 'the universe' does not rotate relative to
local inertial frames inside the void.

2) In an expanding universe,
the matter density as well as the rotational perturbation decrease as a function of time; hence
the influence of the matter on the inertial frames diminishes so that in this limiting case the dragging
inside the void becomes perfect.

\bigskip
$\bullet $ {\it The case of closed universes}
\bigskip

When the universe is closed, $\bar\Omega_{\Sigma}^m$ is positive for all times $\eta_0$ and
for all comoving positions of the shell $\chi_0$; it never vanishes (Fig.\ref{fig4}) - 
there is no perfect dragging in closed universes. Since $\bar\Omega_{\Sigma}^m$ does not depend on the choice of the function
$f(r)$, this result holds for any initial angular momentum distribution of the cosmic dust outside the shell. 
Thus, the influence of the shell on the inertial frames inside the void is always balanced by
the cosmic dust moving to conserve the zero total angular momentum of the universe.  

\bigskip
In both cases, the fact that the whole matter in the universe (the shell and the cosmic dust)
is not observed inside the void at any finite time to (slowly) rotate in a preferred direction 
is in full agreement with Mach's observation. We thus conclude that  
inertial frames inside the void are just in this sense determined by the 'average motion of distant galaxies'.
  

\section*{Acknowledgements}

T.D. would like to thank Guillaume Faye for his help with the numeric part of the paper. We thank Joseph Katz for his
critical as well as friendly reading of the manuscript and dedicate him this work for his 70th birthday. 
J.B. and T.D. acknowledge the support from the grants J13/98:113200004, GACR 202/99/0261 and
GAUK 114/2000. J.B. also acknowledges the kind hospitality of the DARC in Meudon and the Raymond and
Beverly Sackler Distinguished Visiting Fellowship at the Institute of Astronomy in Cambridge. 


\appendix

\section{Extrinsic curvature of the shell}

In the regions inside and outside the void,
the independent tangent vectors at the shell are
\bea&e^{\mu ~in}_t=(1,\dot a l_0,0,0)\quad,\quad e^{\mu ~out}_t=(1,0,0,0),\nonumber\\ 
    & e^{\mu ~in,out}_{\theta}=(0,0,1,0)\quad,\quad e^{\mu ~in,out}_{\phi}=(0,0,0,1),\nonumber\eea
where $\mu$ stands for $(t, r, \theta, \phi)$.
The normal vectors to the shell $n_{\mu}^{in}$ and $n_{\mu}^{out}$,
both pointing inside the shell,
are determined by $(e^{\mu}n_{\mu})^{in,out}=0$, and are normalized to unity:
$$n_{\mu}^{in}=\alpha(t)(\dot al_0,-1,0,0) \quad,\quad n_{\mu}^{out}={a(t)\over \sqrt{1-kl_0^2}}(0,-1,0,0).$$
The extrinsic curvature is defined by
$$K_{ab}^{in,out}\equiv -e_a^{\mu} e_b^{\nu }\nabla_{\mu} n_{\nu}|^{in,out},$$
where the indices $(a,b)$ stand for $(t,\theta,\phi)$ and are raised and lowered by the induced metric 
on the shell (\ref{9}). 
$\nabla_{\mu}$ is the covariant derivative associated with the metric (\ref{2}) or (\ref{5}). 

When the metric is given by (\ref{2}), the non-zero components of the extrinsic curvature of $\Sigma$,
calculated at first order in $\omega_0(t)$, are equal to 
\bea
K^{in}_{tt}&=&-{l_0\ddot a\over\sqrt{1+l_0^2\dot a^2}}\quad,\quad K^{in}_{\theta\theta}=l_0
a\sqrt{1+l_0^2\dot a^2}
\quad,\quad K^{in}_{\phi\phi}=K^{in}_{\theta\theta}\sin^2\theta,\nonumber\\
K^{in}_{t\phi}&=&-l_0 a\omega_0\sqrt{1+l_0^2\dot a^2}\,\sin^2\theta.
\lbl{12}\eea

Similarly the non-zero components of the extrinsic
curvature of
$\Sigma$, when computed with the metric (\ref{5}), are 
\bea
K^{out}_{tt}&=&0\quad,\quad K^{out}_{\theta\theta}=l_0 a\sqrt{1-kl_0^2}
\quad,\quad K^{out}_{\phi\phi}=K^{out}_{\theta\theta}\sin^2\theta,\nonumber\\
K^{out}_{t\phi}&=&-l_0
a\sqrt{1-kl_0^2}\left({1\over2}l_0{\partial\omega\over\partial
r}|_0+\omega_0\right)\,\sin^2\theta
,\lbl{13}\eea 
where $\partial\omega/\partial r|_0\equiv\partial\omega(t,r)/\partial r|_{r=l_{0}}.$

The integration of Einstein's equations across the shell gives the stress-energy tensor ${\cal T}_{ab}$ 
of the shell as [9]
\be -8\pi G{\cal T}_{ab}=\hat K_{ab}-\gamma_{ab}\hat K^c_c,\lbl{14}\ee
$\hat K_{ab}\equiv K^{out}_{ab}-K^{in}_{ab}$
and $\gamma_{ab}$ are the coefficients of the induced metric (\ref{9}).

\section{The metric perturbation {\Large $\omega$}}

Einstein's equations relate the rotational perturbation $\omega(t,r)$ and the dust angular velocity
$\Omega(t,r)$. They can be easily found in the 
the literature (see e.g. [13]), recalling that $\omega(t,r)$ is Bardeen's vector metric perturbation in the
Poisson gauge. In the notations of [13] they read
\be aD_{(k}\dot{\bar B}_{l)}+2\dot aD_{(k}\bar B_{l)}=0\quad,\quad{1\over2}D_iD^i\bar B_k+k\bar B_k=
8\pi Ga^2(\rho+p)(\bar B_k+\bar v_k),\lbl{23}\ee 
where $(k,l)$ stand for $(r,\theta,\phi)$, 
$D_k$ is the covariant derivative associated with the spatial part of the metric (\ref{5}), $\bar
B_k=(0,0,-a\omega r^2\sin^2\theta)$, and $\bar v_k=(0,0,a\Omega r^2\sin^2\theta)$. The first equation (\ref{23}) implies
\be 3\dot a{\partial\omega\over\partial r}+a{\partial\dot\omega\over\partial r}=0,\lbl{24}\ee
the general solution of which is
\be \omega(t,r)=-{1\over a^3}(\beta(r)+g(t)).\lbl{BB}\ee 
The second equation (\ref{23}) translates as:
\be (1-kr^2){\partial^2\omega\over\partial r^2}+(4-5kr^2){1\over r}{\partial\omega\over\partial r}=
-16\pi Ga^2(\rho+p)(\Omega-\omega).\lbl{26}\ee
This can be integrated and, using (\ref{20}) and (\ref{BB}), written in terms of the constant of motion $j_\Sigma$ as
\be
{d\beta\over dr}={6G\over r^4\sqrt{1-kr^2}}(j_\Sigma+j(r)),\lbl{27a}
\ee
where $j(r)$ is the angular momentum of the cosmic dust defined by (\ref{22}).
Further integration of the prevoius equation implies finally
\be
\beta(r)=-{2G\sqrt{1-kr^2}\over r^3}(1+2kr^2)j_\Sigma+6G\int_{l_0}^r{j(r^{\prime})\over
r^{\prime 4}\sqrt{1-kr^{\prime 2}}}\,dr^{\prime}.
\ee

\section{Angular velocity of a distant star measured by inertial observers inside the void}

In order to find $\bar\Omega_{star}^{m}$, we need to compute the trajectories of two photons emitted by
the star, and propagating first in the perturbed FLRW universe 
outside the shell and then in the void inside the shell, and arriving at the centre of the void.

Consider thus a star comoving with the cosmic dust at some radius $\chi_*$ 
(where $\chi$ is as usualy introduced by $r=\chi, \sinh\chi, \sin\chi$ for $k=0,-1,+1$, respectively), 
emitting photons
radially inwards and for the sake of simplicity in the equatorial plane. The trajectory of a photon inside
the shell, as described in the inertial frame ${\cal S}_{in}$, is given by the null geodesic
\be \bar r=\bar t_{rec}^P-\bar t\quad ;\quad\bar\phi=\bar\phi_{rec}^P\quad ;\quad \theta=\pi/2
\quad\hbox{for}\quad \bar r\leq al_0,  \lbl{A9}\ee 
where $\bar t_{rec}^P$ and $\bar\phi_{rec}^P$ are constants and where
$P$ stands for the photon $1$ or $2$. These photons started from the shell at times $\bar t_0^P$ 
defined by $a(\bar t_0^P)l_0=\bar
t_{rec}^P-\bar t_0^P$. Recalling that  the coordinate $\bar\phi$ is related to the coordinate $\phi$ outside
the void by $d\bar\phi=d\phi-\omega_0(t)dt$ and that the proper time in
${\cal S}_{in}$ is related to the time in the outside coordinate grid by $d\bar t=\alpha(t)dt$, 
introducing also the conformal time $\eta$ ($dt=a(\eta)d\eta$), we have 
\bea
\delta\bar t_{rec}&=&[a(\alpha+l_0\dot a)]_0\delta\eta_0\quad \hbox{where}\quad\delta\bar
t_{rec}\equiv \bar t_{rec}^2-\bar t_{rec}^1,\nonumber\\
\delta\bar\phi_{rec}&=&\delta\phi_0-[\omega a]_0\delta\eta_0\quad
\hbox{where}\quad\delta\bar\phi_{rec}\equiv\bar\phi_{rec}^2-\bar\phi_{rec}^1,
\lbl{A10}\eea 
which relate
$\delta\bar t_{rec}$ to $\delta\eta_0$ (the  interval, in the conformal time outside the shell, between the 
arrival times of
the two photons on the shell) and $\delta\bar\phi_{rec}$ to $\delta\phi_0$ (the angular deviation of the two photons
on the shell in the outside coordinate grid), and where the subscript '0' means that the quantities are evaluated at the
time the photons reach the shell.

Outside the shell the ingoing photons propagate, at zeroth order, along the radial null geodesics:
$ d (\bar k^0 a^2)/d\lambda=0$, $d\bar k^r/d\lambda+2a^{-1}da/d\eta\, \bar k^0\bar k^r +(kr)/(1-kr^2)\,(\bar k^r)^2=0$,
$\bar k^{\theta}=\bar k^{\phi}=0$, where
$\lambda$ is an affine parameter and the bar denotes the zeroth order null geodesics in the FLRW unperturbed universe.
The non-zero components of the null wave vector are
$\bar k^0\equiv d\eta/d\lambda=(1/a^2)$ and $\bar k^r\equiv d r/d\lambda=-(1/a^2)\sqrt{1-kr^2}$.
The coordinates $\chi(\eta)$ and $\theta(\eta)$ along the photons are thus given by
\be \chi-\chi_0 = \eta_0^P-\eta\quad,\quad\theta=\pi/2, \lbl{A11}\ee
$\eta_0^P$ is the conformal time at which the photon $P$ reaches the shell at $\chi_0$, $\delta\eta_0 = \eta^2_0-
\eta^1_0$.
The photons are emitted at $\eta=\eta_*$ and $\eta=\eta_*+\delta\eta_*$ by the star placed at 
$\chi=\chi_*$, so that we see from (\ref{A11}) that simply
\be \delta\eta_0=\delta\eta_*\lbl{A12},\ee 
which relates $\delta\eta_0$ to $\delta\eta_*$.

Now at first order, the wave vector of the light propagating in the {\it perturbed} spacetime outside the
shell is  $k^{\mu}=(\bar k^0,\bar k^r,0,k^{\phi})$, $k^\phi\equiv d\phi/d\lambda$ solves the  null geodesic equation
up to first order in the perturbations: 
$${dk^{\phi}\over d\lambda}-\left\{\dot a a\omega+{\partial\over\partial\eta}(a\omega)\right\}(\bar k^0)^2+
2\dot a\bar k^0 k^{\phi}-{1\over r^2}{\partial\over\partial r}\left\{r^2 a \omega\right\}\bar k^0\bar k^r+$$
\be +{2\over r}\bar k^r k^{\phi}+{1\over{1-kr^2}}\left\{\dot a a \omega\right\}(\bar k^r)^2=0,
\lbl{A13}\ee 
where $\omega(t,r)$ is given by (\ref{32}) for $k=0,-1$ and by (\ref{37}) for $k=+1$.
Equation (\ref{A13}) has a simple solution $k^{\phi}=\omega/ a$, where
$\omega$ can now be expressed only as a function of conformal time 
since the trajectory of the photon is given at leading order by (\ref{A11}):
$\omega(\eta,\chi(\eta))=\omega(\eta,\chi_0+\eta_0^P -\eta)$. 
Hence, using $d\eta/d\lambda=\bar k^0=1/a^2$, the $\phi(\eta)$ coordinates of the
trajectories of the photons can be expressed as
\be \phi-\phi_0^P=\int_{\eta_0^P}^{\eta}a(\eta^{\prime})\omega(\eta^{\prime},\chi_0+\eta_0^P -\eta^{\prime})
d\eta^{\prime},\lbl{A14}\ee 
where $\phi_0^P$ are their angles at which they reach the shell, $\delta\phi_0 = \phi^2_0-\phi^1_0$.
From (\ref{A14}), taking into account that $\omega=-a^{-3}(\eta)(\beta(\chi)+g(\eta))$ (Eq.(\ref{25})), we obtain
\be \delta\phi_*-\delta\phi_0=\delta\eta_0\int_{\eta_0^1}^{\eta_*}
\left[\beta(\chi_0+\eta_0^1 -\eta)\times \left({2\over a^3}{da\over d\eta}\right)-
{d\over d\eta}\left({g\over a^2} \right)\right]d\eta,\lbl{A15}\ee
which relates $\delta\phi_0$ to $\delta\eta_*$ and $\delta\phi_*$ (the angular
deviation of the two photons at the time of emission in the outside coordinate grid) - again the subscript '*' means
that the quantities are evaluated at the time of emission. Integrating by parts and changing the variable of
integration $\eta$ to $\chi$,  (\ref{A15}) can be written as
\be \delta\phi_*-\delta\phi_0=\delta\eta_*\left([a\omega]_*-[a\omega]_0-\int_{\chi_0}^{\chi_*}{d(a^3\omega)\over
d\chi}{1\over a^2}d\chi\right),\lbl{A16}\ee
where $a$ is now a function of $(\chi_0+\eta_0 -\chi)$, $\eta_0$ stands for $\eta^1_0$; the index 1 denoting the first
ray is dropped out here and in the following.

In order to relate $\delta\phi_*$ to $\delta\eta_*$ , we use the fact that the star is comoving with the cosmic
dust so that
\be \delta\phi_*=[a\Omega]_*\delta\eta_*\lbl{A17},\ee 
where $\Omega$ is the angular velocity of cosmic matter  introduced in (\ref{21}) and related to
the rotational perturbation $\omega$ by (\ref{31}) for $k=-1,0$ and by (\ref{36}) for $k=+1$.  

Putting the results (\ref{A8}), (\ref{A10}), (\ref{A12}), (\ref{A16}) and (\ref{A17}) together, we finally arrive at
\be \bar\Omega_{star}^{m}={1\over[a(\alpha+l_0\dot a)]_0}\left([a\Omega]_*-[a\omega]_*+
\int_{\chi_0}^{\chi_*}{d(a^3\omega)\over d\chi}{1\over a^2}d\chi\right),\ee 
where we recall that $\alpha=\sqrt{1+l_0^2\dot{a}^2}$.



\begin{thebibliography}{22}
\bibitem{[1]}  G.F. Smoot et al. Astrophy.J. {\bf 371}, L1 (1991)

\bibitem{[2]}  S. Vadas, Phys. Rev. D {\bf 48} (1993) 

\bibitem{[3]}  H. Thirring, Phys. Z. {\bf 19}, 33 (1918); Phys. Z. {\bf 22}, 29 (1921)

\bibitem{[4]}  D. Brill, J.M. Cohen, Phys. Rev. {\bf 143}, 1011 (1966)

\bibitem{[5]}  L. Lindblom, D.R. Brill, Phys. Rev. D {\bf 10}, 3151 (1974)

\bibitem{[6]}  J. Katz, D. Lynden-Bell, J. Bi\v c\'ak, Class. Quantum Grav. {\bf 15}, 1 (1998)

\bibitem{[7]}  C. Klein, Class. Quantum Grav. {\bf 10}, 1619 (1993)

\bibitem{[8]}  S. M. Lewis, General Relativity and Gravitation Vol. {\bf 12}, No. {\bf 11}, 917 (1980)

\bibitem{[9]}  W. Israel, Nuovo Cimento B {\bf 44}, 1 (1966); errata {\bf 48}, 463 (1967)

\bibitem{[10]}  J. M. Bardeen, Phys. Rev. D {\bf 22}, 1882 (1980)

\bibitem{[11]}  H. Bondi, J. Samuel, gr-gc/9607009 (1996)

\bibitem{[12]}  D. Lynden-Bell, J. Katz, J. Bi\v c\'ak, MNRAS {\bf 272}, 150; errata {\bf 277}, 1600 (1995)

\bibitem{[13]}  V.F. Mukhanov, F.A. Feldman, R.H. Brandenberger, Phys. Rep. {\bf 215}, 203 (1992) 

\bibitem{[14]}  C. Klein, Class. Quantum Grav. {\bf 11}, 1539 (1994)

\bibitem{[15]}  C. Barrab\`es, W. Israel, Phys. Rev. D {\bf 43}, 1129 (1991)

\bibitem{[16]}  J. Barbour, H. Pfister eds., Mach's principle : from Newton's bucket to Quantum Gravity, Birkh\"auser, 
Basel (1995)

\end{thebibliography}
\end{document}